\documentclass[twocolumn, prl, showpacs]{revtex4-1}
\usepackage{graphicx}
\usepackage{dcolumn}
\usepackage{bm}
\usepackage[table]{xcolor} 

\begin{document}

\title{Nonstoichiometric doping and Bi antisite defect in single crystal Bi$_2$Se$_3$}

\author{F.-T. Huang$^1$}
\author{M.-W. Chu$^1$}
\author{H. H. Kung$^2$}
\author{W. L. Lee$^2$}
\author{R. Sankar$^1$}
\author{S.-C. Liou$^1$}
\author{K. K. Wu$^{1,3}$}
\author{Y. K. Kuo$^{3}$}
\author{F. C. Chou$^{1,4,5}$}
 \email{fcchou@ntu.edu.tw}
\affiliation{
$^1$Center for Condensed Matter Sciences, National Taiwan University, Taipei 10617, Taiwan}
\affiliation{
$^2$Institute of Physics, Academia Sinica, Taipei 11529, Taiwan}
\affiliation{
$^3$Department of Physics, National Dong Hwa University, Hualien 97401, Taiwan}
\affiliation{
$^4$National Synchrotron Radiation Research Center, Hsinchu 30076, Taiwan}
\affiliation{
$^5$Center for Emerging Material and Advanced Devices, National Taiwan University, Taipei 10617, Taiwan}

\date{\today}

\begin{abstract}
We studied the defects of Bi$_2$Se$_3$ generated from Bridgman growth of stoichiometric and nonstoichiometric self-fluxes. Growth habit, lattice size, and transport properties are strongly affected by the types of defect generated.  Major defect types of Bi$_{Se}$ antisite and partial Bi$_2$-layer intercalation are identified through combined studies of direct atomic-scale imaging with scanning transmission electron microscopy (STEM) in conjunction with energy-dispersive X-ray spectroscopy (STEM-EDX), X-ray diffraction, and Hall effect measurements.  We propose a consistent explanation to the origin of defect type, growth morphology, and transport property.

\end{abstract}



\maketitle

\begin{figure}
\begin{center}
\includegraphics[width=3.3in]{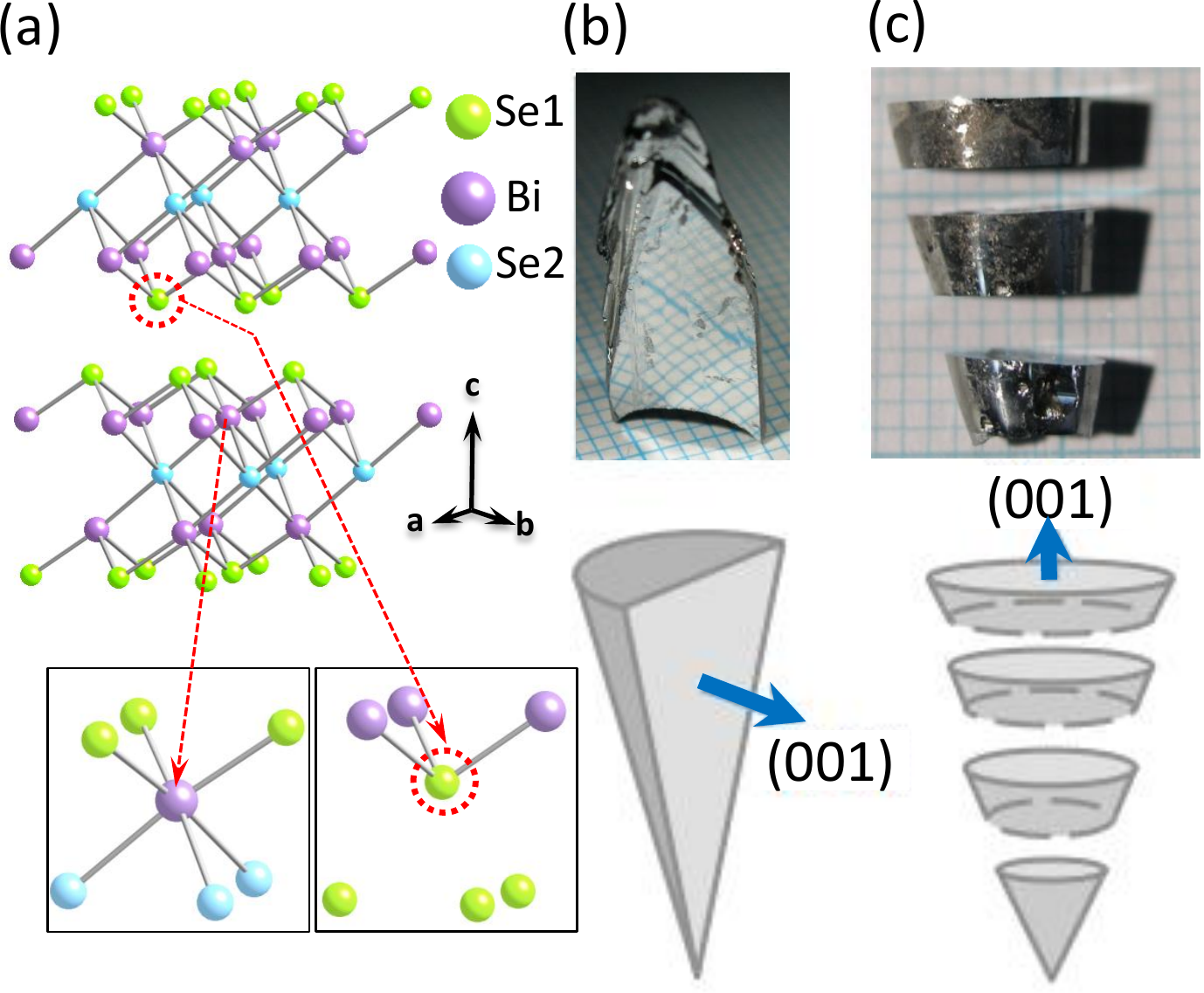}
\end{center}
\caption{\label{fig:fig-crystal}(Color online) (a) Crystal structure of Bi$_2$Se$_3$ can be described with space group R$\overline{3}$m, where quintuple layer is the building block with Se2(blue) in the middle and Se1(green) near the van der Waals gap. The inset shows the BiSe$_6$ octahedron and the Se1 (or Bi$_{Se1}$ antisite in dashed circle) environment. (b) Crystals grown using Bi:Se=2:3 and 2:(3+x) of stoichiometric and Se-rich fluxes, with (001) plane index mostly perpendicular to the growth direction. (c) Crystals grown using Bi:Se=(2+x):3 of Bi-rich flux show (001) plane index along the growth direction.}
\end{figure}

Bi$_2$Se$_3$ is the second-generation topological insulator with a nearly idealized single Dirac cone.\cite{Hasan2010}  Bi$_2$Se$_3$ is also the sister compound of Bi$_2$Te$_3$ that has been demonstrated to show high thermoelectric figure-of-merit (ZT) for commercial applications.\cite{Scherrer1995} Both Bi$_2$Se$_3$ and Bi$_2$Te$_3$ samples show nearly identical crystal structure composed of Bi-(Se,Te) quintuple layers with a van der Waals gap between Se(Te)1, as shown in Fig.~\ref{fig:fig-crystal}(a).  
While Bi$_2$Te$_3$ has been shown with n- or p-type conduction through fine-tuning of nonstoichiometry of Bi/Te ratio, i.e., to dope the system by generating antisite defects of Bi$_{Te}$ (p-type) or Te$_{Bi}$ (n-type),\cite{Fleurial1988} all Bi$_2$Se$_3$ samples reported since 1960 showed n-type semiconducting behavior without exception, and the lowest n-type carrier concentration reported is at the order of 10$^{16}$-10$^{17}$ cm$^{-3}$.\cite{Butch2010}  

Topological insulator is expected to be a band insulator with surface conduction and spin-polarized surface state.\cite{Hasan2010} When the Fermi level is higher in the conduction band as a result of intrinsic n-type doping, it becomes difficult on data interpretation for results containing contributions from both the bulk and the surface.  Angle resolved photoemission spectroscopy (ARPES) studies require n-type doping to map out the shape of the Dirac cone for Bi$_2$Se$_3$, but transport property measurement faces more challenge on distinguishing conductivities from the bulk and the surface.  Additional hole doping methods through Ca substitution and NO$_2$ oxidation have been applied to provide the necessary fine-tuning of Fermi level near the Dirac cone.\cite{Wray2010, Hsieh2009}  It is desirable to have complete carrier type and level control for the topological insulator Bi$_2$Se$_3$ at the stage of crystal growth.

In an effort to grow Bi$_2$Se$_3$ single crystals of controlled carrier type and level for surface-sensitive studies, and to clarify the nature of defects  responsible for the persistent n-type character, we have studied single crystals grown using the Bridgman technique with intentionally generated defects through non-stoichiometric self-fluxes.  With the integrated studies of direct defect observation through STEM-EDX imaging, X-ray diffraction lattice size analysis, and transport property investigation, we were able to find reasonable correlations between the observed physical properties and the distribution of defects in the form of either Bi$_{Se}$ antisite or partially intercalated Bi$_2$-layer patches.

Single crystals studied in this work were grown using the vertical Bridgman method. The initial compounds were high purity bismuth (99.999$\%$) and selenium (99.999$\%$) powder mixed thoroughly in an argon-filled glove box. Three groups of sample were prepared in different Bi:Se molar ratios, including Bi:Se molar ratios of Bi:Se=2:3, (2+x):3, and 2:(3+x) with x=0.05-0.2.  The mixed powder were sealed in quartz tubes under vacuum of about 3$\times$10$^{-2}$ torr after multiple argon gas purging cycles, pre-reacted at 650$^\circ$C for 18 hours in a box furnace, and furnace-cooled to room temperature.  Single crystals were grown with a vertical Bridgman furnace starting from the pre-reacted powder and vacuum-sealed in quartz tubes of 10 cm long and 1.6 cm inner diameter.  The temperature profile of the Bridgman furnace used for the whole series was maintained at 850$^\circ$C-650$^\circ$C in 25 cm. Initial complete melting was achieved at 850$^\circ$C for 24 hours to ensure complete reaction and mixing. The temperature gradient of 1$^\circ$C/cm was programmed around the solidification point near 705$^\circ$C, and the quartz tube was then slowly lowered into the cooling zone at a rate of $\sim$0.5 mm/h. Crystal structure was examined by X-ray diffraction with in-house powder X-ray diffractometer (Bruker D8). Transport properties were examined using a standard four-probe lock-in technique with magnetic field up to 2 Tesla and temperature down to 2 K. High-angle annular dark-field (HAADF) imaging and chemical mapping at atomic-column resolution were performed, with JEOL-2100F microscope equipped with a probe Cs corrector in conjunction with EDX as detailed previously.\cite{Chu2010}. Chemical analysis was performed using Electron probe microanalysis (EPMA) on pieces separated from the as-grown cone-shaped solidified boule. It is interesting to note that the Se-rich flux growth shows growth morphology mostly, but not always, with (001) plane index perpendicular to the growth direction, as shown in Fig.~\ref{fig:fig-crystal}(b). On the other hand, the growth from Bi-rich flux has (001) plane index parallel to the growth direction always and can be cleaved easily into parallel discs, as shown in Fig.~\ref{fig:fig-crystal}(c).

\begin{figure}
\begin{center}
\includegraphics[width=3.5in]{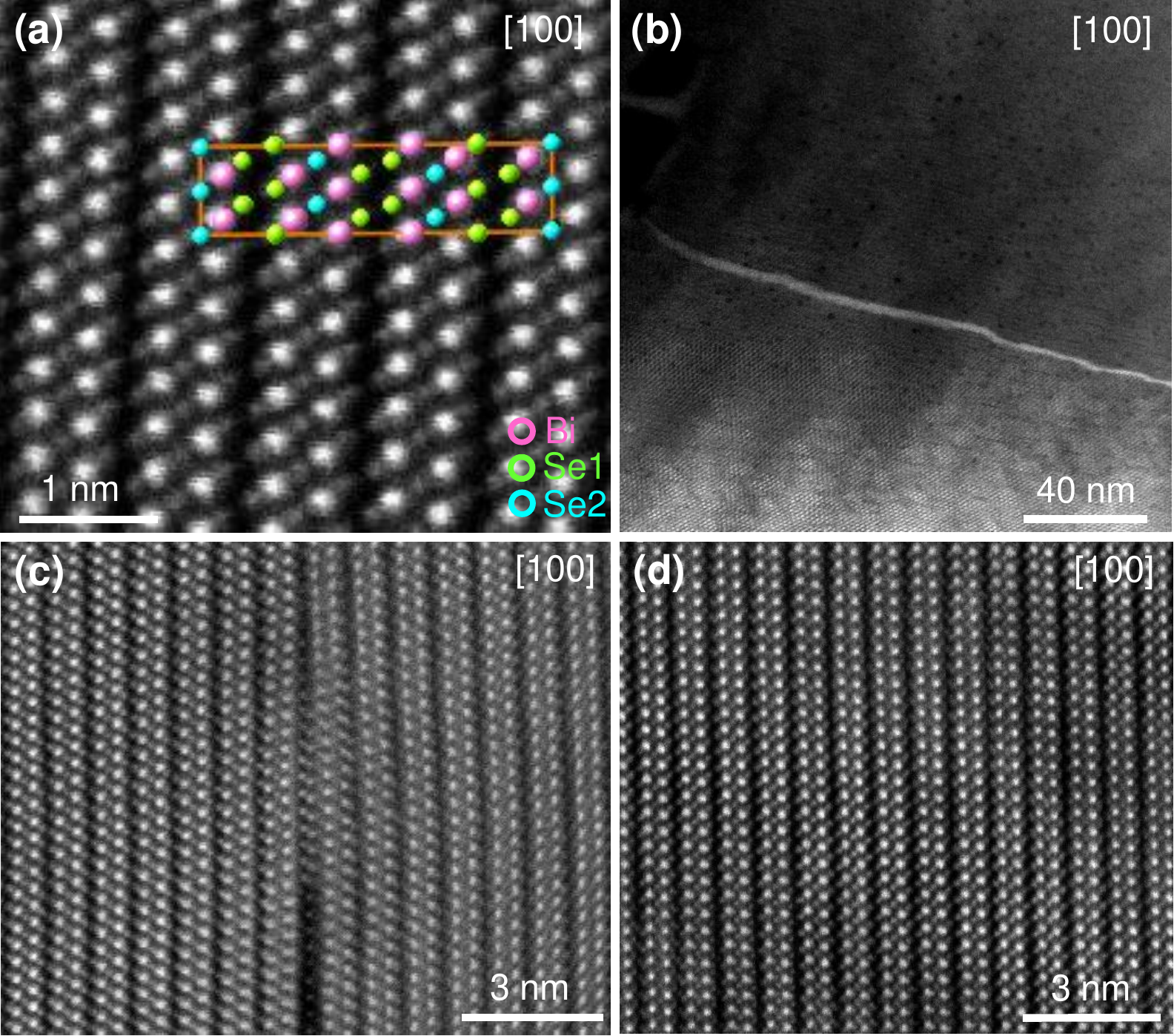}
\end{center}
\caption{\label{fig:fig-defects}(Color online) STEM-HAADF images of Bi$_2$Se$_3$. (a) High resolution image shows that the quintuple of Bi$_2$Se$_3$ consists of two bright spots (Bi) and three fainter spots (Se). (b) For the Bi-rich flux growth, high density of Bi can be found in the grain boundary as shown by the thick and bright lines in low magnification. (c) Low density of Bi$_2$-layer intercalated patches can always be found in the van der Waals gap for the as-grown crystals, whether using Bi- or Se-rich flux. The intercalated Bi$_2$-layer patches would deform the quintuple layer locally by opening up the van der Waals gap. (d) Large area of perfectly ordered quintuple layers without intercalated Bi$_2$-layer patches induced deformation can be found in samples after 300$^\circ$C annealing.}
\end{figure}

STEM-HAADF images for samples selected from different initial growth flux ratios and positions are summarized in Fig.~\ref{fig:fig-defects}. Fig.~\ref{fig:fig-defects}(a) can be viewed as the [100] projection sequenced as quintuple Se1-Bi-Se2-Bi-Se1, which shows two bright spots corresponding to the heavier Bi atoms and the three weaker ones corresponding to the Se atoms as expected, as well as the van der Waals gaps in between the quintuple layers.  For crystals grown from Bi-rich flux in lower magnification, as shown in Fig.~\ref{fig:fig-defects}(b), bright domain boundaries can be found and verified to be Bi, using STEM-EDX characterizations.  As expected for the Bi-rich flux growth, the excess Bi may  aggregate into grain boundaries during cooling, most probably due to the low melting point of Bi (271$^\circ$C), while crystal solidification occurs near 705$^\circ$C. 
Moreover, low density of Bi$_2$-layer patches were found in the van der Waals gap sporadically, for both the Bi-rich and Se-rich flux grown crystals, as shown in Fig.~\ref{fig:fig-defects}(c).  To accommodate the intercalated Bi$_2$-layer patches, the quintuple units near the enlarged van der Waals gap are locally distorted.  Interestingly, large areas of structure free from imperfections could only be found after 300$^\circ$C annealing for both the Se-rich and Bi-rich grown crystals, as shown in Fig.~\ref{fig:fig-defects}(d), which indicates that both Bi excess in the grain boundaries and the intercalated Bi$_2$-layer patches had been removed accordingly.   

\begin{figure}
\begin{center}
\includegraphics[width=3.5in]{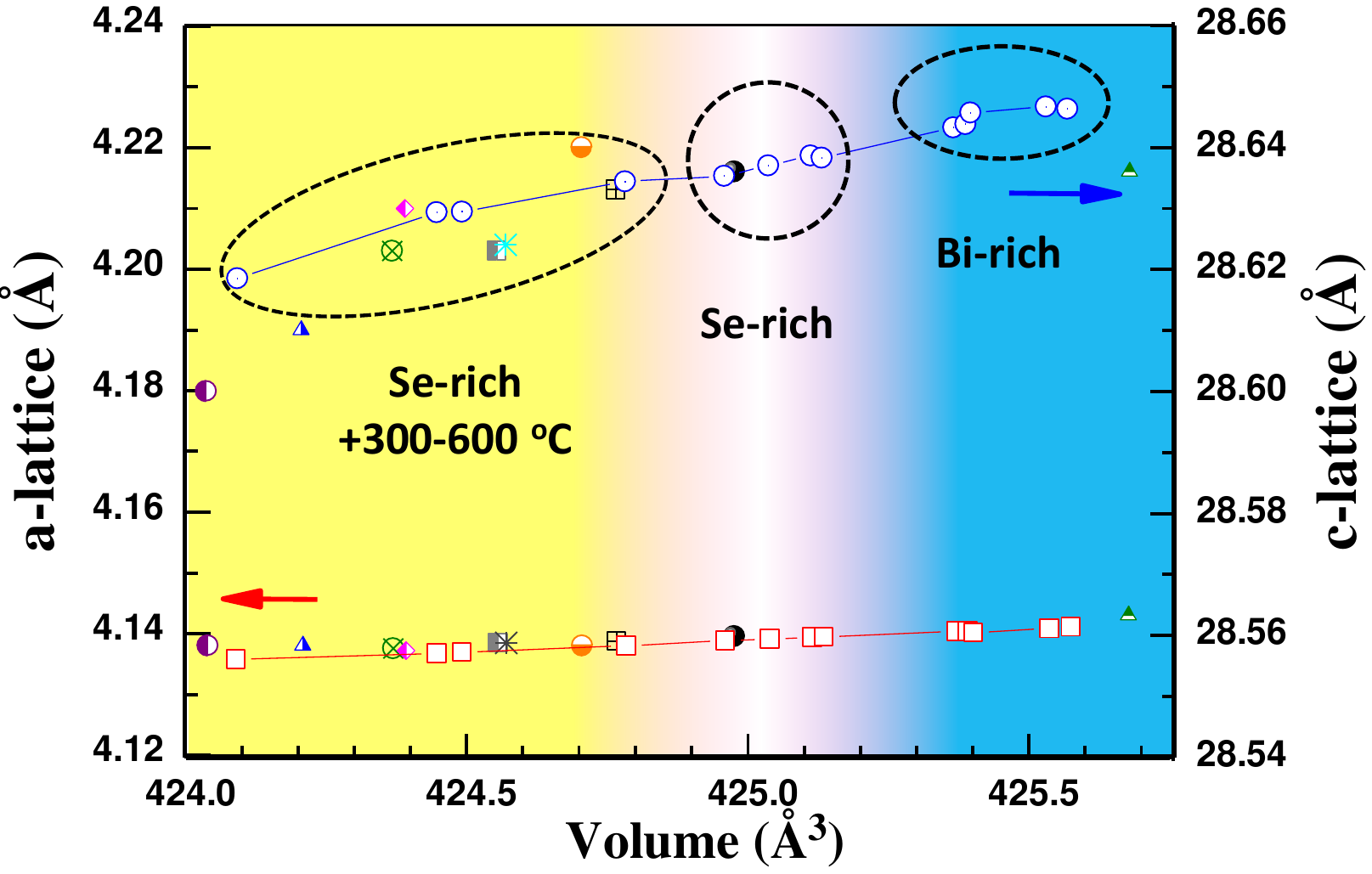}
\end{center}
\caption{\label{fig:fig-lattice}(Color online) Lattice parameters vs. lattice volume of all crystals studied, where \textbf{c}-axes are in empty circles and \textbf{a}-axes are in empty squares. All studied samples are clustered in groups of Bi-rich flux growth before annealing (Bi-rich), Se-rich flux growth before annealing (Se-rich), and Se-rich plus post annealing at 300-600$^\circ$C (Se-rich+300-600$^\circ$C), in an increasing trend as indicated by the connected experimental data points.  A list of Bi$_2$Se$_3$ lattice parameters reported in the literature since 1960 are shown in various solid symbols.\cite{all}}
\end{figure}

Lattice parameters of \textbf{c}-axis have been examined using X-ray diffraction for pieces selected from different sections of the grown boule. Additional annealing process has been applied at 300$^\circ$C to 600$^\circ$C in an evacuated sealed tube for part of the as-grown crystals. Lattice parameters are summarized in Fig.~\ref{fig:fig-lattice} and plotted in the \textbf{a}- and \textbf{c}-axes against lattice volume. Interestingly, all studied samples are clustered in groups of Bi-rich flux growth before annealing (Bi-rich), Se-rich flux growth before annealing (Se-rich), and Se-rich plus 300-600$^\circ$C post annealing (Se-rich+300-600$^\circ$C), in a trend of increasing volume and \textbf{c}-axis.  In addition, a collection of lattice parameters of Bi$_2$Se$_3$ published since 1960 are displayed together and found falling nicely along the same increasing trend.\cite{all}  It is clear that  many of the inconsistent results published previously could be due to the subtle differences in the nonstoichiometry of Bi and Se. 

We may postulate the reason why crystals from Bi-rich flux growth have longer \textbf{c}-axes is due to the larger amount of Bi intercalated into the van der Waals gaps, mostly in the form of randomly distributed patches of neutral metal Bi$_2$-layer as revealed in the STEM picture of Fig.~\ref{fig:fig-defects}(c), but not enough to form the metastable phases of staged (Bi$_2$)$_m$(Bi$_2$Se$_3$)$_n$ at the high solidification point near 705$^\circ$C.\cite{Lind2003}  It should be noted that the staged phases beyond Bi$_2$Se$_3$ can only be prepared at the much lower temperature range of $\sim$400-560$^\circ$C as intercalated metastable compounds.\cite{Okamoto1994}  In particular, the average lattice volume increase should be viewed as randomly distributed intercalated Bi$_2$-layer patches without changing the original R$\bar{3}$m symmetry, especially when the local dilation of van der Waals gap occurs only near regions with intercalated Bi$_2$-layer patches.  In addition, 300$^\circ$C annealing reduces the lattice size significantly, which indicates that a small amount of intercalated Bi can be removed with a temperature above the melting point of Bi effectively, as supported also by the creation of large area of defect-free structure after 300$^\circ$C annealing shown in Fig.~\ref{fig:fig-defects}(d).  On the other hand, high annealing temperature up to 600$^\circ$C starts to decompose the sample partially, presumably as a result of too many Se vacancies created in an evacuated environment, which has also been verified by thermogravimetric analysis under argon flow with the following X-ray structure analysis. 

\begin{figure}
\begin{center}
\includegraphics[width=3.3in]{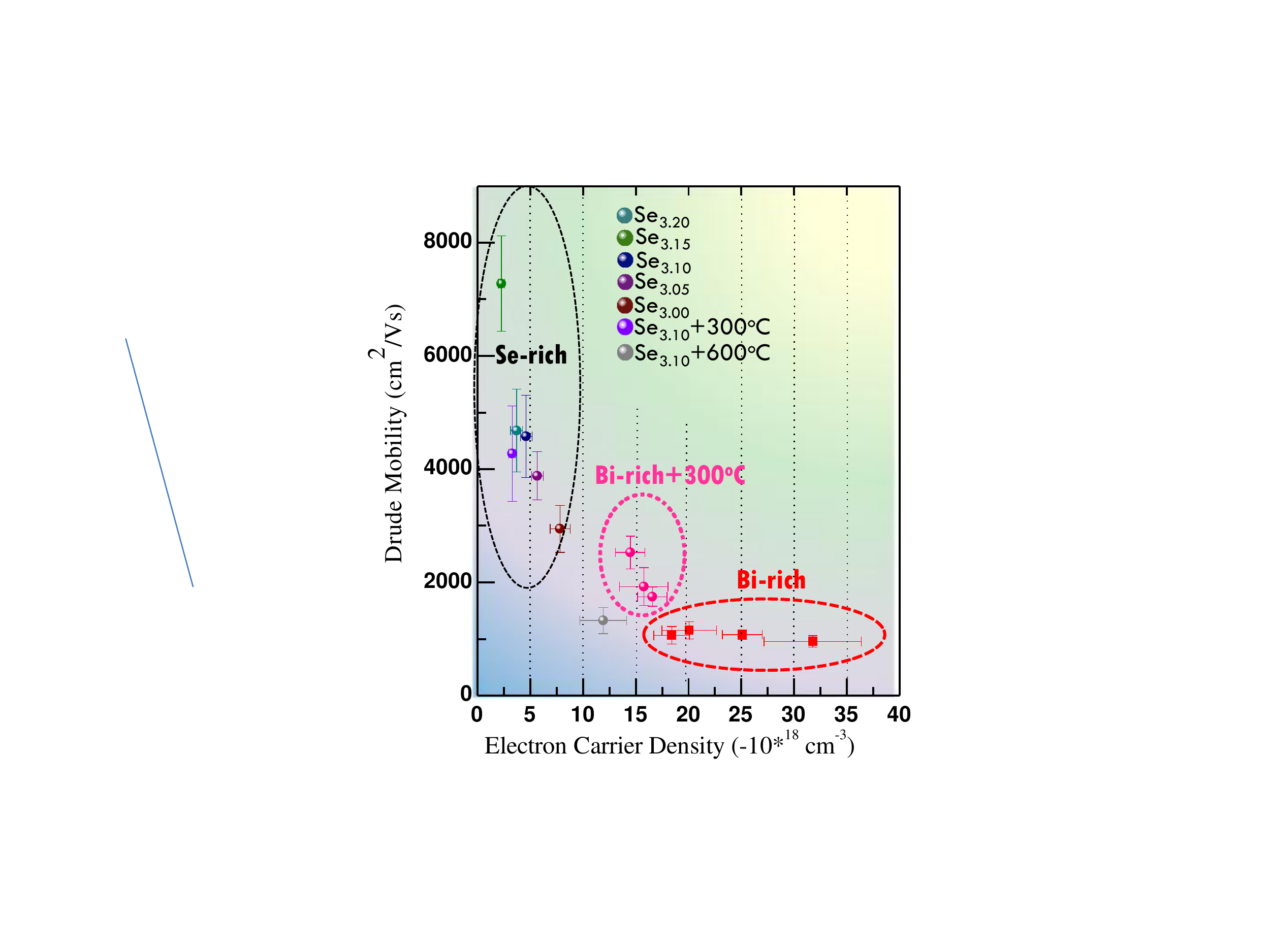}
\end{center}
\caption{\label{fig:fig-mobility}(Color online) Drude mobility ($\mu$) vs. carrier concentration ($n$) for all samples studied are shown; all are n-type. Crystals grown from Bi-rich flux (Bi-rich) have the highest carrier concentration but the lowest mobility, and post 300$^\circ$C annealing (Bi-rich+300$^\circ$C) reduces the carrier concentrations to the same level of $\sim$-15$\times$10$^{18}$.  Crystals from stoichiometric and Se-rich flux growth (Se-rich) have the lowest carrier concentration and highest mobility, and the 300$^\circ$C post annealing (Se$_{3.10}$+300$^\circ$C) shows no significant impact on modifying the carrier concentration and mobility.}
\end{figure}

All single crystal samples show metallic behavior down to about 20 K (not shown, similar to those published earlier)\cite{Sugama2001}, below which the resistivity $\rho$ becomes nearly temperature-independent, as expected in the impurity scattering regime. Fig.~\ref{fig:fig-mobility} summarizes the relationship between carrier concentration $n$ and mobility $\mu$ for all samples studied. The carrier concentration $n$ was determined from Hall effect measurement at 2 K, and the corresponding Drude mobility $\mu$ was estimated using Drude formula $1/\rho = n e \mu$. For the stoichiometric and Se-rich flux growth, samples are clustered at the low carrier concentration $\sim-5\times10^{18}$ /cm$^{3}$ and high mobility regime. On the other hand, crystals grown from Bi-rich flux have the highest carrier concentrations and the lowest mobility.  In addition, 300$^\circ$C annealing has no significant impact on the already low carrier concentration from Se-rich flux growth, while the carrier concentrations for Bi-rich flux growth crystals are reduced to a common lower level near $\sim-15\times10^{18}$ /cm$^{3}$.   

Based on evidences from partial decomposition after 600$^\circ$C annealing in an evacuated tube, thermogravimetric analysis, and lattice size analysis, Se vacancy must be limited to a low threshold before the occurrence of decomposition. This is because Se has a vapor point of 685$^\circ$C, below the solidification point of Bi$_2$Se$_3$ near 705$^\circ$C, which leads to the occurrence of solidification at the liquid-gas phase boundary in the Bridgman growth.  It is nearly impossible to maintain a perfect Bi-Se stoichiometry upon Bridgman solidification near 705$^\circ$C, especially for the relatively weakly bonded Se1 at the van der Waals gap.  We conclude that the Se vacancy is impossible to avoid under the current Bridgman growth condition of limited Se vapor pressure.  The Se vacancy level must be determined mainly by the Se partial pressure during Bridgman growth, instead of the initial Bi to Se ratio, not to mention the fact that the liquid Bi in the flux prefers to intercalate into the van der Waals gap after the Bi$_2$Se$_3$ solidification, instead of forming Bi-Se quintuple layer of high Se vacancy level.  Although lacking rigorous quantitative evidence, current results seem to agree with the commonly accepted assumption that n-type carriers are generated from Se-vacancy formation, i.e., each Se vacancy donates at most two itinerant electrons as Se$_{Se}\rightleftharpoons$V$^{\cdot\cdot}_{Se}$+Se$_{(g)}$+2e$^-$ in the Kroger-Vink notation of various doping efficiency.\cite{Drasar2010} 

\begin{figure}
\begin{center}
\includegraphics[width=3.5in]{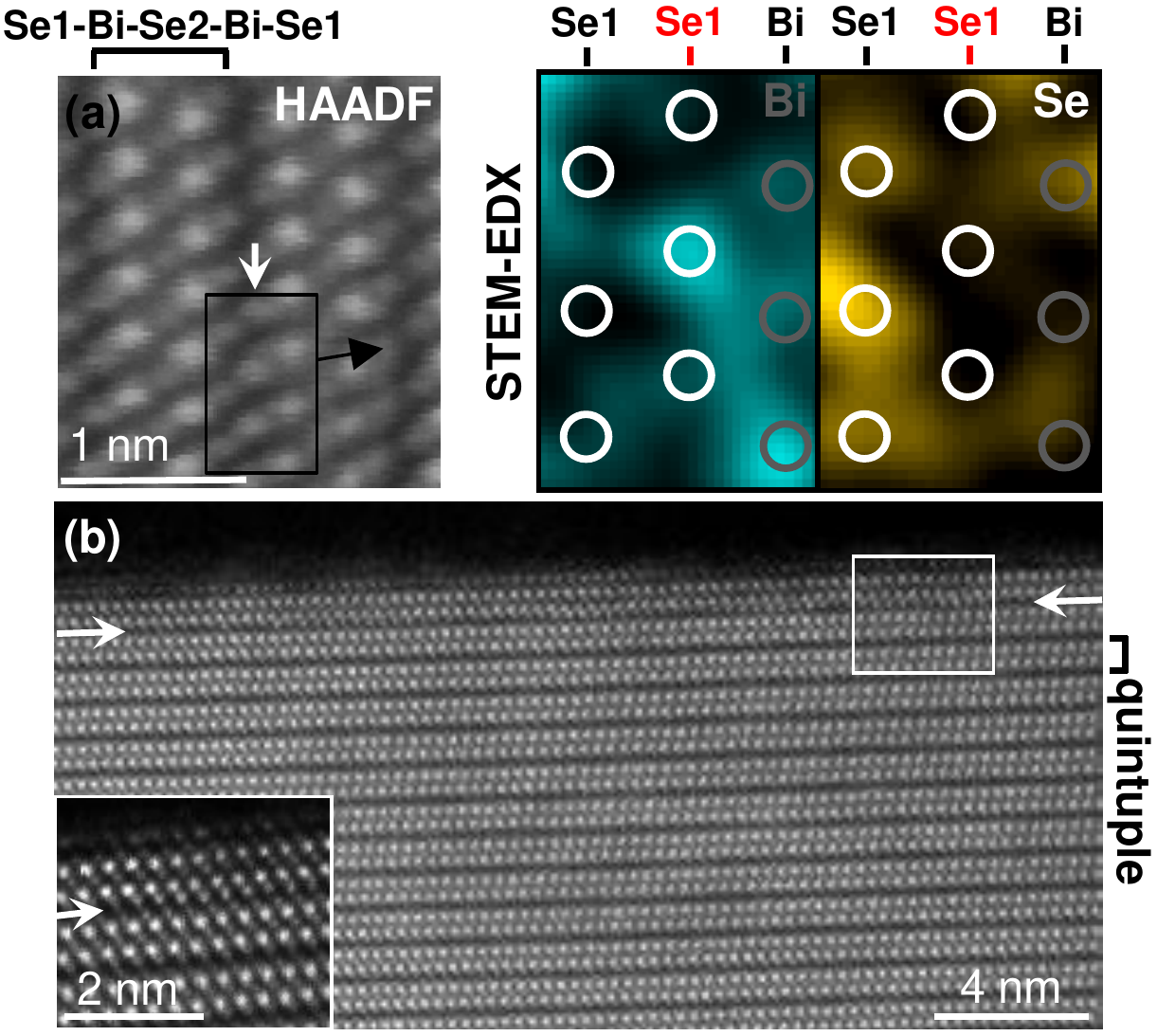}
\end{center}
\caption{\label{fig:fig-antisite}(Color online) (a) STEM-HAADF image of a bulk region with thickness $\sim$25 nm along the incident beam direction.  Se1 column pointed out by the white arrow shows an anomalously enhanced imaging contrast.   Further STEM-EDX chemical mappings of the corresponding boxed region are enlarged in the right panels (cyan, Bi map; yellow, Se map).  The gray (white) circles in the mapping represent the Bi (Se) columns.  The enhanced contrasts of the Se1 columns (signified by the red Se1 text) are shown to be correlated with the presence of Bi on the sites, i.e., the formation of Bi$_{Se}$ antisite defects. (b) The enhanced Se1 contrasts are more regularly observed at the edge of the same specimen as indicated by the white arrows, with corresponding enhanced blow-up of the white boxed region in the inset in lower left corner.}
\end{figure}

In addition to the finding of intercalated Bi$_2$-layer patches, as revealed in Fig.~\ref{fig:fig-defects}(c), we also observed Bi$_{Se}$ antisite defects in the bulk of $\sim$25 nm thickness along the incident beam direction, as shown in Fig.~\ref{fig:fig-antisite}, using atomically-resolved STEM-EDX mapping.  We notice that some columns of supposed Se1 arrays near the van der Waals gap show unexpected enhanced contrast, as shown in Fig.~\ref{fig:fig-antisite}(a).  The atomically resolved chemical mapping by STEM-EDX shown in the right panels of Fig.~\ref{fig:fig-antisite}(a) reveals that the enhanced contrast at the Se1 site actually corresponds to Bi dominant character, which provides a direct proof of partial substitution of Bi to the Se1 site, i.e., the Bi$_{Se}$ antisite defects formation.  The observed Se1 columns with an enhanced contrast towards the sample edge (Fig.~\ref{fig:fig-antisite}(b)) signify the existence of more antisite defects near the surface layers.  The antisite defects both in the bulk and near the surface region has been a general phenomenon throughout our STEM-EDX characterization of various batches of samples.   

The importance of antisite defect in the tetradymite-type compounds has been discussed in the scope of narrow band gap semiconductor previously,\cite{Drasar2010} yet quantitative evidence is difficult to obtain at the point defect level. 
Considering the number of outer shell electrons for Bi(6s$^2$6p$^3$) and Se(4s$^2$4p$^4$), simple argument borrowed from extrinsic semiconductor  suggests that Bi$_{Se}$ antisite can be an acceptor.  However, it would be a picture too simplified when Bi$_{Se}$ antisite defect is actually formed by the Bi substitution only to the Se1 vacancy site near the van der Waals gap, i.e., Bi$_{Se1}$ antisite, and chemically, the electronegativity difference between Se(2.55) and Bi(2.02) is too large for Bi to accept electrons further from the neighboring Bi/Se atoms. Originally, Se1 uses its four outer shell electrons in 4p$^4$ to bond with the neighboring Bi's and the two 4s$^2$ electrons form lone pairs exposed to the van der Waals gap, as shown in the inset of Fig.~\ref{fig:fig-crystal}(a).\cite{Drabble1958}  The bonding nature of the substituted Bi in Bi$_{Se1}$ antisite becomes unknown in view of a virtual octahedral environment of Bi(Bi$_3$Se1$_3$) as depicted in the inset of Fig.~\ref{fig:fig-crystal}(a). Current experimentally confirmed n-type donor and the direct STEM observation of Bi$_{Se1}$ antisite defect supports a scenario that Bi$_{Se1}$ might be viewed as a pair of V$_{Se1}$ and Bi interstitial, i.e., the Se vacancy acts as electron donor and Bi remains neutral.  This scenario is reasonable, considering that the van der Waals radius of Bi (2.07$\AA$) is very close to the ionic radius of Se$^{2-}$ (VI=1.98$\AA$).\cite{Mantina2009}  Indeed, based on the formation energy calculated using first-principles calculations, Bi$_{Se1}$ antisite defect has been shown to be favorable, compared to the Se vacancy defect with a V$_{Se1}$ character.\cite{Wang2012}

The Bi$_2$-layer intercalation must introduce severe lattice distortion near regions where the van der Waals gap starts to dilate in order to accommodate the excess Bi, as shown in Fig.~\ref{fig:fig-defects}(c).  
Bi$_{Se1}$ antisite may introduce less strain to the system compared to the intercalated Bi in the van der Waals gap, but both can break the inversion symmetry locally, which might be associated with the observed symmetry-forbidden peaks found in Raman spectrum of thin film studies.\cite{Shahil2010, Zhao2011, Cheng2011}  There are relatively more Bi$_{Se1}$ aitisite defects found near the specimen edges than in the bulk from studies of multiple batches of either as-grown or 300$^\circ$C annealed samples, as demonstrated in Fig.~\ref{fig:fig-antisite}(b).  While surface defects could cause severe band bending away from the Dirac point at the surface, current findings based on STEM observation alone require other experimental techniques to verify.  


In summary, we conclude that crystals grown using Se-rich flux for the Bridgman growth method can reduce Se vacancy level significantly; however, it is not possible to prevent Se vacancy formation to achieve p-type doping using excess Se under current the Bridgman growth condition.  Two major types of defects are observed in Bi$_2$Se$_3$ crystals grown using Se-rich or Bi-rich flux: the randomly intercalated neutral Bi$_2$-layer patches in the van der Waals gaps, and the Bi$_{Se1}$ antisite point defects that act presumably as electron donor.  Low temperature annealing of Se-rich flux grown crystal at 300$^\circ$C helps to remove the intercalated Bi$_2$-layer patches effectively and can create a large area of ordered near stoichiometric compound.

FCC acknowledges the support from National Science Council of Taiwan under project number NSC-100-2119-M-002-021 and Academia Sinica under project number AS-100-TP2-A01.


\end{document}